# Field Stability in Direct Modulation of Semiconductor Laser

Hoang Nam Nhat
Faculty of Physics, College of Science, Vietnam National University
334 Nguyen Trai, Thanh Xuan, Hanoi, Vietnam
namnhat@hn.vnn.vn

*Abstract*— We studied the rate equations of direct modulation laser and showed that it may be reduced to the special case when spontaneous carrier decay rate is equal to the photon decay rate. The solution in this case is unique. For the general case, we investigated the vector field of the differential system of the rate equations and pointed out the basic stability problems of this system when the modulation current was required to change.

## I. INTRODUCTION

Recently, several works appeared to show that the direct modulation of semiconductor laser is possible [1-5]. The successful application of direct modulation would bring a general benefit in symplifying the physical equipment and reducing its price. By introducing of an appropriate modulation current, the undesirable features of semiconductor lasers that originated in the nonlinear dynamics such as period doubling cascades, period tripling and chaos might be suppressed. The clearing of the effect so-called "inter-symbol interference", which causes damage to the future signals just by the presenting ones, and which has its physics in the existence of finit relaxation time during switching the optic and the electrical field [6-7], opens the posibility for the high speed transmission of signal using high frequency direct modulation.

Usually the laser operates in the external modulation by applying the changing external voltage. This change affects the optic output $s(t)$ which can be decoded after transmission at the receiver. The direct modulation, instead, involves changing the current input around the bias level. It is principally simpler method but unfortunately it produces the output that depends on internal dynamics of laser, that is on the interplay between optic and electrical field. Neither the characteristics of optic nor of electric component can be known apriori but are measured experimentally.

Up-to-date there are generally two methods for shaping the current inputs: (1) using piece-wise discontinuous constant current [4] and (2) using shaped continuous current [5]. The later one has just appeared on the preprint in Apr 2004 and seemed to bring additional advantage to the method. The operation of direct modulation laser depends, however, heavily on the generating internal noise so the correlation between laser input and output may not be constant during operation, i.e. the laser may not return to its start fixed state after some operating cycles. In our opinion this problem was fundamental and should be investigated more carefully since the natural noise of laser system, which usually increases with operating frequency might damage the output signals when communication speed increased.

In the next Sec. II we review the technique that authors in [5] has used to discuss the direct modulation of laser and in Sec. III present some primary results on the stability of phase portrait of laser rate equation.

## II. RATE EQUATION

### A. Single mode laser dimensionless rate equation

The single-mode laser is a dynamic system driven by two physical variables, the photon density $S(t)$ and the carrier density $N(t)$:

$$\frac{dS}{dt} = -\gamma_c S + \Gamma G(N,S)S$$
$$\frac{dN}{dt} = -\gamma_s N - G(N,S)S + \frac{J_0 + J}{ed} \quad (1)$$

where $J_0$ is the bias current, $J(t)$ the modulation (around the bias), $G(N,S)$ the optical gain, $\Gamma$ the confinement factor, $d$ the active layer thickness. The $\gamma_c$ and $\gamma_s$ are two important physical constants: the photon decay rate ($\gamma_c$) and the spontaneous carrier decay rate ($\gamma_s$).

By choosing some fixed point ($N_0$, $S_0$) with known gain coefficients ($\Gamma G_0 = \gamma_c$, $J_0/ed - \gamma_s N_0 = G_0 S_0$), the $G(N,S)$ may be expanded linearly around this point yielding:

$$G(N,S) = G_0 + G_n(N-N_0) + G_p(S-S_0) \quad (2)$$

By substituting this to the equation (1) and by introducing the dimensionless quantities, the authors in [5] has achieved the equations:

TABLE I.  VARIABLES AND NUMERICAL VALUES

| Symbol [a] | Value | Description |
|---|---|---|
| $\omega_R$ | $2.222\times 10^{10}s^{-1}$ | $\sqrt{\gamma_c\gamma_n+\gamma_s\gamma_p}\approx 1/45 ps$ |
| $\gamma_s$ | $1.458\times 10^{9}s^{-1}$ | spontaneous carrier decay rate |
| $\gamma_n$ | $1.333\times 10^{9}s^{-1}$ | gain variation with carrier density |
| $\gamma_p$ | $2.400\times 10^{9}s^{-1}$ | gain variation with photon density |
| $\gamma_c$ | $3.600\times 10^{11}s^{-1}$ | photon decay rate |
| **Coeficient** | **Value** | **Description** |
| A | 0.108 | $A=\gamma_p/\omega_R$ |
| B | 0.060 | $B=\gamma_n/\omega_R$ |
| C | 0.066 | $C=\gamma_s/\omega_R$ |
| D | 16.2 | $D=\gamma_c/\omega_R$ |
| J | 2/3 | bias current at fixed point |
| A/D | 0.00667 | $A/D=\gamma_p/\gamma_c$ |
| B/D | 0.00370 | $B/D=\gamma_n/\gamma_c$ |
| C/D | 0.00405 | $C/D=\gamma_s/\gamma_c$ |
| B/C | 0.91449 | $B/C=\gamma_n/\gamma_s$ |
| JC | 0.04374 | $JC=J\gamma_s/\omega_R$ |
| B/J | 0.09000 | $B/J=\gamma_n/J\omega_R$ |
| JCA/D | 0.00029 | $JCA/D=J\gamma_s\gamma_p/\gamma_c\omega_R$ |
| JC(1+A/D) | 0.04403 | $JC(1+A/D)=J\gamma_s/\omega_R(1+\gamma_p/\gamma_c)$ |
| $CJ_M^{min}$ | -0.109 | $1+J+J_M^{min}=0 \to J_M^{min}=-1-J$ |
| $CJ_M^{max}$ | 0.284 | $1+J+J_M^{max}=6 \to J_M^{max}=5-J$ |

a. Symbols definition and values mainly followed [5]

$$\frac{ds}{d\tau}=\frac{\gamma_c}{\omega_R}[g(n,s)-1]s$$
$$\frac{dn}{d\tau}=\frac{\gamma_s}{\omega_R}[J+J_M-n-Jg(n,s)s] \quad (3)$$

where $g(n,s)$ is:

$$g(n,s)=1+\frac{\gamma_n}{J\gamma_s}n-\frac{\gamma_p}{\gamma_c}(s-1) \quad (4)$$

and the dimensionless quantities are defined as followed: output photon density $s(t)=S(t)/S_0$, input carrier density $n(t)=N(t)/N_0-1$, optical gain $g(n,s)=G(N,S)/G_0$, gain variation with carrier density $\gamma_n=G_nS_0$, gain variation with photon density $\gamma_p=-\Gamma G_pS_0$, natural relaxation oscillation angular frequency $\omega_R=\sqrt{\gamma_c\gamma_n+\gamma_s\gamma_p}$, dimensionless time $\tau=t\omega_R$ and current $J+J_M=[J_0+J(t)]/[\gamma_sN_0ed-1]$.

We adopt this notion about the dimensionless quantities since it is useful in reducing the differential system (3) furthermore to obtainning the simplified dependence of $J_M$ on $n(t)$ and $s(t)$ as described in the next section.

### B. Modulation current

By inserting (4) into (3), first for $ds/d\tau$ and then for $dn/d\tau$ (see the symbols and constants listed in Table I), we obtain:

$$s'=\frac{ds}{d\tau}=-As^2+\frac{B}{J}sn+As$$
$$n'=\frac{dn}{d\tau}=-Bsn-Cn+JC\frac{A}{D}s^2-JC\left(1+\frac{A}{D}\right)s+CJ+CJ_M \quad (5)$$

This Hamiltonian system has been used to construct the vector field featured in Fig. 1 whose co-variant vector $\vec{c}=(n',s')$ is directly given by (5).

Substitute $B\frac{C}{D}sn-B\frac{C}{D}sn$ into the second equation and observe that $-B\frac{C}{D}sn+JAs^2\frac{C}{D}-JAs\frac{C}{D}$ is just $-\frac{C}{D}s'$ we can arrive at:

$$J_M=\left(\frac{n'}{C}+n\right)+J\left(\frac{s'}{D}+s-1\right)-\frac{B}{C}\left(\frac{C}{D}-1\right)sn \quad (6)$$

This equation expresses the symmetrical contribution of the carrier ($n$) and the photon ($s$) density to $J_M$, especially for the case when $C=D$, i.e. when the spontaneous carrier decay rate $\gamma_s$ is equal to the photon decay rate $\gamma_c$. The symmetry strikes even more if one observes that $s'=(s-1)'$ so one may substitute $X$ for $(s-1)$, $Y$ for $n$, and rewrite the relation (6) as ($C=D=K$):

$$J_M=\left(\frac{Y'}{K}+Y\right)+J\left(\frac{X'}{K}+X\right) \quad (7)$$

A little algebra will reduce this further to the one dimensional system:

$$J_M=Z+\frac{1}{K}Z' \quad (8)$$

by assuming a general variable $Z=Y+JX=n+J(s-1)$. This equation is easily solvable with a general solution of form:

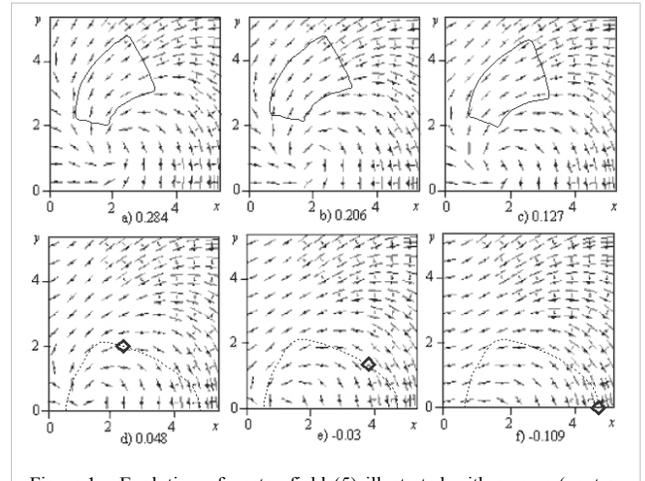

Figure 1. Evolution of vector field (5) illustrated with arrows (contra-varianta) and stacks (co-variants) under the variation of dimensionless modulation current $CJ_M$ in the range (-0.109, 0.284). The units in both axes ($y=n$, $x=s$) are arbitrary. The values of $CJ_M$ must satisfy the condition $0<1+J+J_M<6$. Other parameters were as listed in Table I. The polygon in each frame shows the closed geodesics that conserved the variants, i.e. it was drawn so that its parts did either cross the contra-variants or the co-variants at the same value. The small squares show the evolution of the fixed point ($s_0$, $n_0$) during the development of $J_M$.

$$Z = e^{-Kt} \int K J_M(t) e^{Kt} dt \quad (9)$$

These two equations (8) and (9) totally determine dynamics of laser with the property $\gamma_s = \gamma_c$. By giving either $Z$ or $J_M$ the counterpart is determined uniquely. We found this theoretical case extremely interesting since it set no limit to the communication speed. Whether or not $\gamma_s = \gamma_c$ is physically possible, it is worth attention and further consideration.

We now leave this case aside and move on the discussion about the perturbation of Hamiltonian system (5) under variation of $J_M$ and other factors.

### III. VECTOR FIELD OF THE RATE EQUATION

#### A. Dynamics of vector field under variation of $J_M$

It is illustrative to consider (5) as a two parameter vector field $(n,s)$ where the perturbation of it due to $J_M$ or other factors may be directly shown on the graph. Fig. 1 shows up 6 phases in the development of field (5) with the axis defined as $y=n$, $x=s$ and $CJ_M$ varies from lowest -0.109 to highest value 0.284. The limits for $J_M$ were calculated so that the condition $0<1+J+J_M<6$ holds as required in [5]. This pictures, being independent to the function $s(t)$ and $n(t)$, resemble the storm forming process with its center moving up-left. It is evident that the closed geodesics that conserve the variants do not carry the fixed point $(s_0, n_0)$ to itself during the variation of $J_M$. So the system $(J_M, s_0, n_0)$ would return to itself if and only if the $J_M$ would reset itself to the original value. We found this condition very critical to (5) in contrast to [5] where the authors appeared to relax from this. Indeed, according to [5] the $(s_0, n_0)$ might evolute to itself even when $J_M$ would modulate.

#### B. Stability of vector field under perbutation of other factors

Since the evolution of $J_M$ was not a unique factor what drived the field (5), the effect of which could be separable from the other only if the change caused by $J_M$ was enough larger than that by other factors. In case these factors could drive the field out from the state just given by some $J_M$, to some new states associated with the "still not existed" $J_M$ so the output $s(t)$ would not be correct. To illustrate this situation, Fig.2 shows the correlation between $J_M$ and the constants A, B, C, D in creating the same vector field within a given restricted area of the state-space. This correlation may be explained for the constant A as followed: first, let $J_M$ changes from a starting value $J_M^0$, associated with $A^0$, to some value $J_M^{(t)}$ at time $t$. We then evaluate an appropriate value of $A^{(t)}$ satisfying the condition:

$$\begin{aligned} s\cdot\left(J_M^{(t)}, A^0\right) - s\cdot\left(J_M^0, A^{(t)}\right) \approx 0 \\ n\cdot\left(J_M^{(t)}, A^0\right) - n\cdot\left(J_M^0, A^{(t)}\right) \approx 0 \end{aligned} \quad (10)$$

The meaning of this correlation is obvious: since the geometrically identical (static) vector fields produce the algebraically identical (static) solutions then if the variation of a constant, either A, B, C or D can "reset" the field so the $s(t)$ and $n(t)$ will not correspond to the correct modulation current $J_M^{(t)}$. This consequently leads to the corrupted open trajectories in the state-space and the system loses its knowledge about the origin, the "low" and "high" current.

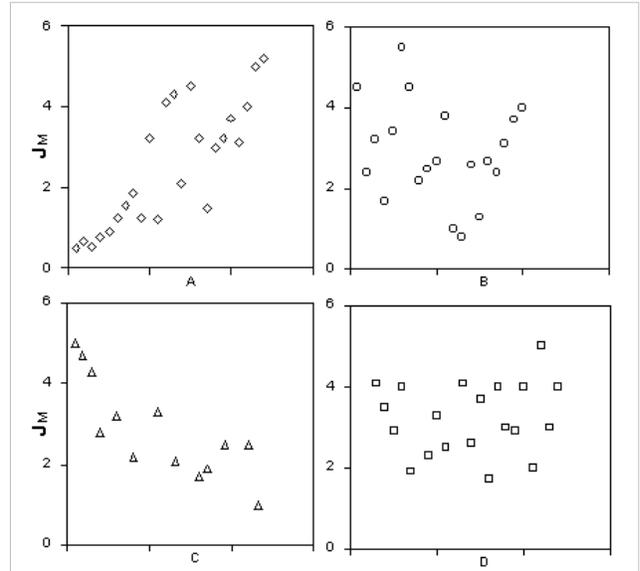

Figure 2. Correlation between $J_M$ and the constants A, B, C, D according to the criteria given in the equation (10) (creating the same vector field). The unit is arbitrary. It is not clear any analytical dependence between these perturbative factors, on which the analytical solutions of the differential system (5) heavily depend on.

#### C. Geodesics that conserve variants

To allow only the $J_M$ curves that bear the appropriate $s(t)$ and $n(t)$ that move a fixed point $(s_0, n_0)$ along the geodesics conserving the variants may be the additional criteria to stabilize the differential system (5) againt the perturbative factors. This is equivalent to leaving the fixed point to float on the geodesics. Certainly we face the problem of how to define such geodesics when conditions change. We leave this for consideration in the future.

### IV. CONCLUSION

The differential system (5) would be stable and solvable uniquely for one special case when the spontaneous carrier decay rate was equal to the photon decay rate. Beyond this the stability of the system is questionable and it appears that there were various correlations between the modulation current $J_M$ and the other physical constants in the way they could damage the state-space portrait and change the position of the fixed point on this state-space. The high speed application of the direct modulation technique in communication relies a lot on the stability of its characteristics and this problem should be investigated in more details. We found some conclusion on the applicability of this technique (e.g. in [5]) more-less hurried.

ACKNOWLEDGMENT

This work has been financially supported by the Vietnam National University HN Research Program QT-04-05 and the Hanoi University of Sciences Research Program TN-04-06. We express our sincere thanks to the Management Units of these programs, especially to the Department of Science and Technology, HUS for their supports and advices.